\documentclass[%
 reprint,
 amsmath,amssymb,
 aps,
pra,
]{revtex4-2}

\usepackage{graphicx}
\usepackage{dcolumn}
\usepackage{bm}

\begin{document}

\preprint{APS/123-QED}

\title{Towards dense optical communications with partially coherent light:\\I. Bimodal Stokes communications}

\author{Ayman F. Abouraddy}
\email{raddy@creol.ucf.edu}
\author{Bahaa E. A. Saleh}%
\affiliation{CREOL, The College of Optics \& Photonics, University of Central Florida, Orlando, Florida 32816, USA}%

\date{\today}

\begin{abstract}
In conventional optical multiplexing, $N_{\mathrm{f}}$~orthogonal spectral or spatial modes multiplex $N_{\mathrm{f}}$ independent data streams transmitted over an optical channel (e.g., a multimode optical fiber supporting $N_{\mathrm{f}}$~guided modes). We introduce here a new optical communications scheme that we call `Stokes communications' (S-Comm), whereby $N_{\mathrm{D}}=N_{\mathrm{f}}^{2}-1$ independent data streams are multiplexed into $N_{\mathrm{f}}$~optical modes, potentially yielding a dramatic increase in the channel capacity. Rather than modulating the amplitude of each mode to encode a data stream, we make use of multimode partially coherent light and modulate each generalized Stokes parameter in the $N_{\mathrm{f}}\times N_{\mathrm{f}}$ Hermitian coherence matrix associated with the field. We present here designs for the S-Comm transmitter and receiver making use of bimodal fields ($N_{\mathrm{f}}=2$) in which $N_{\mathrm{D}}=3$ data streams are encoded via on-off keying and multiplexed in two modes. We describe the main blocks in the S-Comm transmitter: coherence modulators that encode a data stream in the degree of coherence of a bimode (a two-mode partially coherent field), Stokes encoders that assign each data stream to a prescribed Stokes parameter, and Stokes multiplexers that combine bimodes into physical modes. We also describe the reverse operations implemented in the S-Comm receiver to recover the data streams. Significantly higher channel capacity is expected for a larger number of physical modes, which is facilitated by the amenability of the presented designs to on-chip integration. These results suggest that the long-appreciated but unexploited higher information-carrying capacity of partially coherent light can help realize dense optical communications.
\end{abstract}

\maketitle


\section{Introduction}

Consider an optical field spanned by $N_{\mathrm{f}}$~physical modes, which we take to be fixed, deterministic, and stable field configurations that are maintained by the optical system of interest \cite{Abouraddy26AOP}. Such a scenario arises in myriad settings in optics. An example is the field in a multimode optical fiber (MMF) supporting $N_{\mathrm{f}}$~guided modes \cite{Winzer13chapter,Richardson13NP,Li14AOP,Willner15AOP}. The profiles of the MMF modes are dictated solely by the boundary conditions. Consequently, any field in the MMF is constrained to be a superposition of these modes \cite{SalehBook07}. Partial coherence can arise in such settings when the complex modal coefficients in the field superposition are random variables, whereupon the field is described by an $N_{\mathrm{f}}\times N_{\mathrm{f}}$ Hermitian, positive semi-definite coherence matrix $\mathbf{G}$ that tabulates the statistical correlations between pairs of modes \cite{Gori06OL,Kagalwala13NP,Abouraddy14OL,Kagalwala15SR,Abouraddy17OE} -- a configuration we called `structured coherence' \cite{Abouraddy26AOP,Abouraddy26OPN}. It is well-known that the coherence matrix $\mathbf{G}$ is uniquely identified by $N_{\mathrm{D}}=N_{\mathrm{f}}^{2}-1$ real parameters \cite{Waller12NP}, in contrast to the $\mathcal{O}(N_{\mathrm{f}})$ real parameters needed to identify its coherent counterpart, in which the complex modal coefficients are deterministic rather than random. The coherence matrix $\mathbf{G}$~for a partially coherent field can thus be viewed as informationally richer than the description of a coherent optical field sharing the same modal basis \cite{Waller12NP}. Although this richness -- that is manifest in the larger number of tunable free parameters -- increases the complexity of field synthesis, one may nevertheless consider utilizing each real parameter identifying $\mathbf{G}$~as an independent information carrier. This suggests the prospect of dense optical communications exceeding conventional bounds.

Consider again an MMF supporting $N_{\mathrm{f}}$~modes. Each mode can encode a data stream via on-off keying, and the fiber can thus multiplex $N_{\mathrm{f}}$~modes at the same wavelength by assigning each data stream to a different guided mode -- a scheme known as spatial-mode multiplexing \cite{Winzer13chapter,Richardson13NP,Li14AOP,Willner15AOP}. However, if a partially coherent field described by the $N_{\mathrm{f}}\times N_{\mathrm{f}}$ coherence matrix~$\mathbf{G}$, which is identified by $N_{\mathrm{D}}\sim N_{\mathrm{f}}^{2}$ Stokes parameters, is transmitted over this MMF, one can envision multiplexing $\mathcal{O}(N_{\mathrm{f}}^{2})$~data streams in the same $N_{\mathrm{f}}$~physical modes. We called such a scheme `Stokes communications' (S-Comm). Realizing this prospect necessitates devising new optical systems that can assign a data stream to each of the $N_{\mathrm{D}}$~generalized Stokes parameters identifying the coherence matrix $\mathbf{G}$ at the transmitter, and then retrieve the data streams at the receiver after traversing the optical channel. To date, such systems have not been forthcoming (see Ref.~\cite{Nardi22OL} for an initial theoretical suggestion). While utilizing optical coherence has been previously investigated for both optical sensing \cite{Brooks85JLT} and communications \cite{Griffin95JLT,Sampson97FiberIO,Kats04IEEETWC,Meijerink07IEEEJSTQE}, the purpose was \textit{not} to yield a higher channel capacity. In general, partially coherent optical fields have not yet been exploited as carriers for dense optical communications.

\begin{figure*}[t!]
\centering
\includegraphics[width=6.5in]{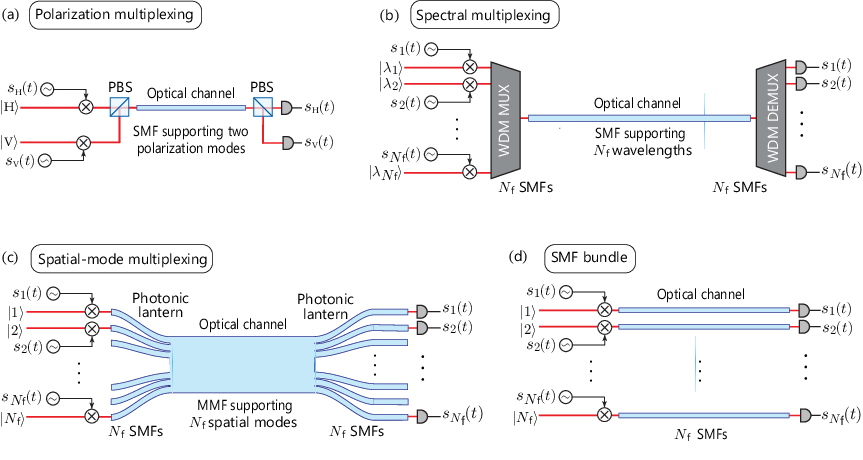}
\caption{Examples of conventional multiplexing. (a) Polarization multiplexing. The data streams $s_{\mathrm{H}}(t)$ and $s_{\mathrm{V}}(t)$ are encoded in $|\mathrm{H}\rangle$ and $|\mathrm{V}\rangle$, respectively; PBS: polarizing beam splitter. (b) Spectral multiplexing. The spectral modes or wavelengths $|\lambda_{j}\rangle$ are modulated by the data streams $s_{j}(t)$ and combined via wavelength-division multiplexing (WDM); MUX: multiplexer, DEMUX: demultiplexer. (c) Spatial-mode multiplexing in an MMF supporting $N_{\mathrm{f}}$~guided modes. Each data stream $s_{j}(t)$ modulates a single mode $|j\rangle$, which are multiplexed into the MMF and subsequently separated via photonic lanterns. (d) A bundle of SMFs. The mode $|j\rangle$ in each SMF is modulated by the data stream $s_{j}(t)$. Strictly speaking, there is no multiplexer or demultiplexer. Each data stream is coupled to one SMF and the outputs are recorded separately.}
\label{Fig:Multiplexers}
\end{figure*}

Here we present an explicit design for an S-Comm transmitter and receiver that enable multiplexing $N_{\mathrm{D}}$ independent data streams in $N_{\mathrm{f}}$ physical modes using partially coherent light. We restrict ourselves here to bimodal optical fields $N_{\mathrm{f}}=2$ and present the generalization of S-Comm to multimode fields $N_{\mathrm{f}}>2$ in an accompanying paper \cite{Abouraddy26JOSAA2}. The design for bimodal S-Comm transmitter involves several new concepts: (1) `bimodes' (two-mode partially coherent fields) \cite{Saleh25book}, each of which enables encoding a single data stream; (2) coherence modulators that encode a data stream into a bimode; (3) Stokes encoders that assign a data stream to a prescribed Stokes parameter of the field; and (4) a Stokes multiplexer that combines the bimodes into physical modes. Moreover, we describe the reverse operations in the S-Comm receiver: (1) Stokes demultiplexing -- whereby the physical modes are separated into bimodes, (2) Stokes decoding in conjunction with (3) differential detection to retrieve the data streams. All of these proposed devices are amenable to on-chip integration and were indeed recently demonstrated in our work on manipulation of multimode partially coherent light in an integrated circuit comprising a mesh of Mach-Zehnder interferometers \cite{Hashemi26arxivTwoMode,Hashemi26arxiv4Modes,Hashemi26LPR3modes,Hashemi26OL}. This is crucial because the multimode S-Comm generalization described in the accompanying paper makes use of the bimodal components as building blocks. We thus lay the groundwork for exploiting multimode partially coherent light in dense optical communications, and add to the emerging swathe of applications of optical information processing in which partially coherent light can outperform coherent light \cite{Peng21P,Liu23PRAppl,Dong24Nature,roques2024Light,Miller25Optica,Liu25LPR,Harling25APLP,Mor26arxiv,Abouraddy26OPN}.

\section{Conventional multiplexing}

\subsection{Physical modes and conventional modulation}

We first describe examples of `physical modes' that are relevant for multiplexing in optical communications. The simplest example is the polarization degree-of-freedom (DoF) that is spanned by two modes denoted $|\mathrm{H}\rangle$ and $|\mathrm{V}\rangle$ \cite{Ip07JLT} (making use of the Dirac notation \cite{Abouraddy26AOP}). A second example is the spectral DoF where individual wavelengths can serve as information carriers \cite{Brackett90IEEEJSAC,Agrawal21Book}. In principle, the number of spectral modes is limited by the optical-channel bandwidth. Finally, the spatial DoF can be used for multiplexing, where orthogonal spatial modes are exploited as information carriers in spatial-mode multiplexing \cite{Winzer13chapter,Richardson13NP}. The number of spatial modes is limited by the numerical aperture, whether in an MMF \cite{Li14AOP}, a free-space link \cite{Willner15AOP}, or an underwater link \cite{Baghdady16OE}. These physical modes correspond mathematically to orthogonal modes of the field. In optical communications, each of these $N_{\mathrm{f}}$~modes is modulated by a single data stream, $s_{j}(t)$, $j=1,\cdots,N_{\mathrm{f}}$. In general, when utilizing on-off keying to modulate the modal amplitude via a conventional optical modulator, the number of independent data streams coincides with the number of physical modes. Starting with an input mode having constant power $I_{\mathrm{in}}(t)=I_{\mathrm{o}}$, its amplitude is modulated by the data stream $s_{j}(t)$ to yield $I_{j}(t)=s_{j}(t)I_{\mathrm{in}}(t)=I_{\mathrm{o}}s_{j}(t)$, where $s_{j}(t)$ is digitized to take on the values~0 or~1.

\begin{figure*}[t!]
\centering
\includegraphics[width=6.5in]{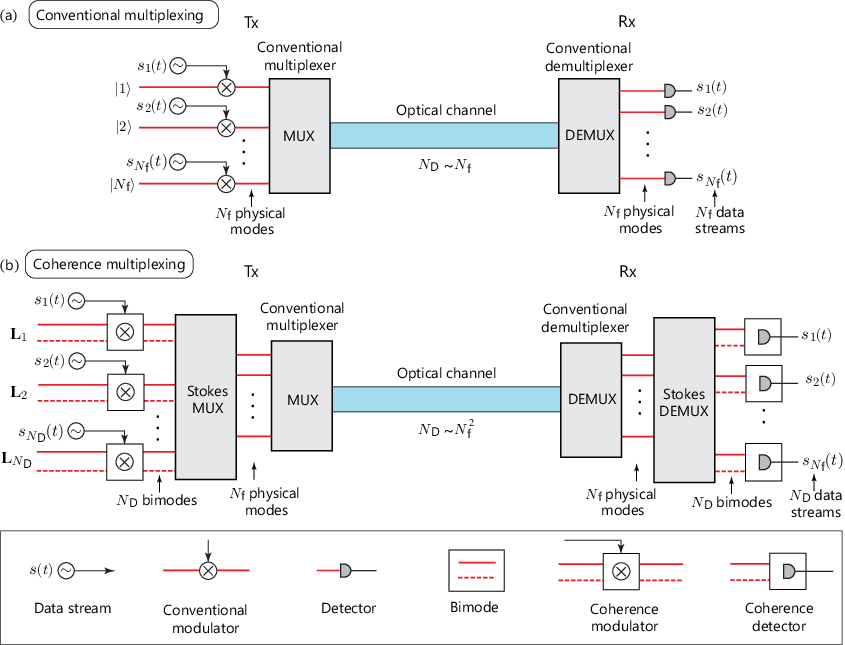}
\caption{Conventional multiplexing versus coherence multiplexing. (a) A generic optical communications scheme exploiting $N_{\mathrm{f}}$ physical modes and making use of a conventional multiplexer (MUX) and demultiplexer (DEMUX). In the transmitter, the $N_{\mathrm{f}}$ modes are each modified by a data stream $s_{j}(t)$, $j=1,\cdots,N_{\mathrm{f}}$, and are then combined by the multiplexer into the optical channel. At the receiver, the demultiplexer separates the $N_{\mathrm{f}}$~physical modes for detection and recovery of the data streams. (b) Envisioned dense optical communications scheme making use of coherence multiplexing. The core of the system is still the conventional multiplexer, optical channel, followed by conventional demultiplexer. However, rather than feeding the multiplexer with $N_{\mathrm{f}}$~modulated physical modes, we first prepare $\sim N_{\mathrm{f}}^{2}$~bimodes (or coherence modes), each of whose degree of coherence is modulated by a data stream. The coherence (or Stokes) multiplexer then combines the $N_{\mathrm{D}}\sim N_{\mathrm{f}}^{2}$~bimodes into $N_{\mathrm{f}}$~physical modes directed to the conventional multiplexer. At the receiver, the $N_{\mathrm{f}}$~physical modes are separated by the conventional demultiplexer, which are in turn separated by the Stokes demultiplexer into $N_{\mathrm{D}}\sim N_{\mathrm{f}}^{2}$~bimodes whose degree of coherence is detected to recover the data streams.}
\label{Fig:Concept}
\end{figure*}

\subsection{Requirements for a conventional multiplexer/demultiplexer}

What facilitates the use of particular DoFs for optical communications is the availability of \textit{multiplexers} and \textit{demultiplexers} operating on these DoFs. A multiplexer combines the modulated modes associated with a particular DoF and delivers the field to an optical channel. Conversely, the field emerging from the optical channel is separated by a demultiplexer into individual modes to recover the data streams. Several criteria are desirable in a multiplexer. (1) \textit{A multiplexer is a fixed device} that combines the $N_{\mathrm{f}}$~physical modes into the optical channel; i.e., no changes are implemented during a bit or from bit to bit. Because optical communications is typically performed at high rates, requiring the multiplexer to perform operations \textit{during} a bit necessitates a process operating at an even higher rate, which is typically unavailable and renders the multiplexer less useful. (2) \textit{A demultiplexer is a multiplexer operated in reverse}, which separates $N_{\mathrm{f}}$~physical modes from the optical channel so that the signals can be recovered independently of each other via high-speed detectors. It is not required to carry out any steps \textit{during} the temporal width of a bit. (3) \textit{A multiplexer is constrained by the number of physical modes.} The set of $N_{\mathrm{f}}$~modes forms a basis -- a maximal set of linearly dependent modes (indeed in our case orthogonal modes). Consequently, the number of multiplexed physical modes is equal to the number of modes for the DoF.

\subsection{Examples of multiplexing DoFs}

In \textit{polarization} multiplexing, the modes $|\mathrm{H}\rangle$ and $|\mathrm{V}\rangle$ are modulated by the data streams $s_{\mathrm{H}}(t)$ and $s_{\mathrm{V}}(t)$, respectively, and then combined into a single spatial mode via a polarizing beam splitter (PBS) serving as a multiplexer [Fig.~\ref{Fig:Multiplexers}(a)]. The field can be coupled to an SMF or free-space link, after which a second PBS demultiplexes the polarization modes, so that detectors can recover the data streams. \textit{Spectral} multiplexing [Fig.~\ref{Fig:Multiplexers}(b)] is heavily used in optical communications, where each wavelength mode $|\lambda_{j}\rangle$ is modulated by a data stream $s_{j}(t)$ and then combined with other wavelengths into a single spatial mode via a wavelength-division multiplexer, with the reverse procedure separating the wavelengths for detection at the receiver.

In \textit{spatial-mode multiplexing}, each data stream modulates a spatial mode, which are multiplexed into an optical channel. If the optical channel is an MMF supporting $N_{\mathrm{f}}$~modes, then each modulated mode is mapped to one of the MMF guided modes via a photonic lantern \cite{LeonSaval10OE,Davenport21OME} or an MPLC \cite{Martinez-Becerril24OE}, with the procedure reversed at the MMF output [Fig.~\ref{Fig:Multiplexers}(c)]. This overall system can also be used in a free-space link. Finally, in a bundle of SMFs, each mode is modulated by a data stream and coupled to one SMF, whose output is detected separately [Fig.~\ref{Fig:Multiplexers}(d)]. Although such a bundle of SMFs is \textit{not} strictly speaking an example of multiplexing, it will be useful once we describe Stokes multiplexing below. 

In these examples, the multiplexer and demultiplexer satisfy the above-listed desiderata: (1) each is a fixed device; (2) the number of data streams multiplexed is equal to the number of physical modes; and (3) the demultiplexer is a multiplexer operated in reverse. The overall communications scheme is illustrated in Fig.~\ref{Fig:Concept}(a). Each of the $N_{\mathrm{f}}$~modes is modulated by a data stream $s_{j}(t)$, $j=1,\cdots,N_{\mathrm{f}}$, and is multiplexed into the optical channel. At the receiver, a demultiplexer separates the physical modes from which the encoded data streams are recovered.

\section{Stokes communications}

The overall envisioned system for Stokes communications (S-Comm) is depicted in Fig.~\ref{Fig:Concept}(b). We still rely on $N_{\mathrm{f}}$~physical modes at the S-Comm transmitter, which are multiplexed into the optical channel via a conventional multiplexer. However, we do \textit{not} modulate the physical modes themselves. Rather, we initially prepare $N_{\mathrm{D}}=N_{\mathrm{f}}^{2}-1$ `coherence modes' and then utilize a new system we refer to as a `coherence modulator', whose function is to modulate the degree of coherence of a single coherence mode according to a digitized data stream. Here, the coherence modes are simply a pair of mutually uncorrelated modes, which we refer to henceforth as `bimodes' \cite{Saleh25book}. Once all $N_{\mathrm{D}}$ bimodes are thus modulated, they are combined via a `Stokes multiplexer' into the $N_{\mathrm{f}}$~physical modes. We thus rely on a three-step procedure: in step~(1), $N_{\mathrm{D}}$~bimodes are prepared and modulated; in step~(2), the modulated bimodes are combined into $N_{\mathrm{f}}$ physical modes; and in step~(3), the $N_{\mathrm{f}}$~physical modes are multiplexed into the optical channel as usual. We have already addressed step~(3) of conventional multiplexing [Fig.~\ref{Fig:Multiplexers}], and we focus here on the first two steps. 

At the S-Comm receiver, the reverse procedure is implemented. First, a conventional demultiplexer separates the $N_{\mathrm{f}}$~modes. Second, a Stokes demultiplexer (a Stokes multiplexer in reverse) separates the $N_{\mathrm{f}}$~modes into $N_{\mathrm{D}}$~bimodes. Finally, a differential detector retrieves the data streams from the bimodes. Crucially, both the Stokes multiplexer and demultiplexer are fixed devices. Such an overall scheme -- if successful -- promises dramatic increase in the communications capacity of an optical channel. We describe here in detail the construction of the major blocks of the S-Comm transmitter and receiver for bimodal fields ($N_{\mathrm{f}}=2$). We describe the extension of this scheme to multimode S-Comm ($N_{\mathrm{f}}>2$) in the accompanying paper \cite{Abouraddy26JOSAA2}. Crucially, the devices introduced here for bimodal S-Comm serve as building blocks for multimode S-Comm.

\subsection{Coherence mode or bimode}

Rather than encoding data streams into the amplitudes of physical modes, we make use instead of a `bimode' whose degree of coherence is modulated by the data stream. We use `bimode' henceforth to refer to both the two-mode field \textit{and} the $2\times2$ coherence matrix representing it, where the coherence matrix is Hermitian, positive semi-definite, and unity trace \cite{Abouraddy26AOP}. The degree of coherence of a bimode is $D=\lambda_{1}-\lambda_{2}$, where $\lambda_{1}$ and $\lambda_{2}$ are the eigenvalues of its coherence matrix.

\begin{figure}[t!]
\centering
\includegraphics[width=8.0cm]{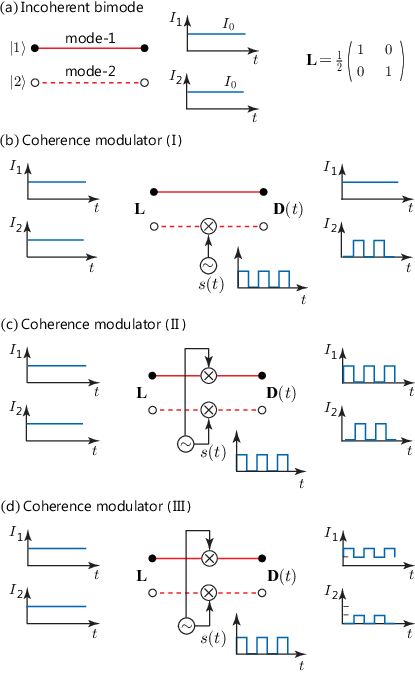}
\caption{(a) Graphical depiction of a bimode described by the coherence matrix $\mathbf{L}$. (b) Design for coherence modulator (I). The input is the bimode $\mathbf{L}$ along with the data stream $s(t)$, and the output is the coherence-modulated bimode $\mathbf{D}(t)$. We plot the modal weights $I_{1}(t)$ and $I_{2}(t)$ at the input and output. (c) Coherence modulator (II), and (d) coherence modulator (III). In both (c) and (d), the bimode produced has the form of $\mathbf{D}(t)$ in Eq.~\ref{eq:Dmatrix}, but $D(t)=\pm1$ in (c), while $D(t)=0,1$ in (d).}
\label{Fig:CoherenceModulation}
\end{figure}

Each bimode initially comprises a pair of mutually uncorrelated modes that do not display interference when superposed. We denote the modes as mode-1 and mode-2, depicted graphically as a pair of lines, solid for mode-1 and dashed for mode-2 [Fig.~\ref{Fig:CoherenceModulation}(a)], and their power levels as $I_{1}(t)$ and $I_{2}(t)$, respectively. For equal power, the bimode is [Fig.~\ref{Fig:CoherenceModulation}(a)]:
\begin{equation}\label{eq:Lmatrix}
\mathbf{L}=\frac{1}{2}\left(\begin{array}{cc}1&0\\0&1\end{array}\right),
\end{equation}
with $I_{1}(t)=I_{2}(t)$, and a degree of coherence $D=0$.

\subsection{Coherence modulator}

We consider a digital data stream $s(t)$ that takes on the values~0 and~1, and examine three different designs for a coherence modulator, such that $s(t)$ modulates the degree of coherence $D(t)$ of a bimode. A coherence modulator transforms the bimode from $\mathbf{L}$ to $\mathbf{D}(t)$, which we refer to as a coherence-modulated bimode.

\subsubsection{Coherence modulator (I)}

In one embodiment [Fig.~\ref{Fig:CoherenceModulation}(b)], the data stream $s(t)$ is sent to a conventional modulator that impacts mode-2 of the bimode, while mode-1 remains unchanged (the two modes can of course be switched). When $s(t)=0$, the resulting coherence-encoded bimode is $\mathbf{D}(t)=\mathbf{L}$, with $D=0$; and when $s(t)=1$, then $\mathbf{D}(t)=\left(\begin{array}{cc}1&0\\0&0\end{array}\right)$ with $D=1$, so that in general $\mathbf{D}(t)=\tfrac{1}{1+s(t)}\left(\begin{array}{cc}1&0\\0&s(t)\end{array}\right)$ with $D(t)=\tfrac{1-s(t)}{1+s(t)}$. Note that the bimode total power changes between the two settings for $s(t)$.

\subsubsection{Coherence modulator (II)}

In the embodiment depicted in Fig.~\ref{Fig:CoherenceModulation}(c), two conventional modulators are utilized, one for mode-1 and the other for mode-2. Both modulators are fed with the data stream $s(t)$ but they work in a complementary fashion: $s(t)=1$ yields $I_{1}=I_{\mathrm{o}}$ and $I_{2}=0$, so that $\mathbf{D}=\left(\begin{array}{cc}1&0\\0&0\end{array}\right)$; and $s(t)=0$ yields $I_{1}=0$ and $I_{2}=I_{\mathrm{o}}$, so that $\mathbf{D}=\left(\begin{array}{cc}0&0\\0&1\end{array}\right)$. The coherence-encoded bimode is:
\begin{equation}\label{eq:Dmatrix}
\mathbf{D}(t)=\frac{1}{2}\left(\begin{array}{cc}1+s'(t)&0\\0&1-s'(t)\end{array}\right),
\end{equation}
with a degree of coherence $D(t)=s'(t)$, where $s'(t)=2s(t)-1$, \textit{and} the total power in the bimode is fixed from bit to bit. However, $D(t)$ here switches between the values 1 and $-1$. In other words, the bimode switches between two orthogonal, coherent field configurations.

\subsubsection{Coherence modulator (III)}

In the third embodiment [Fig.~\ref{Fig:CoherenceModulation}(d)], mode-1 and mode-2 each traverse a conventional modulator fed by the data stream $s(t)$, but with different modulation depths so that $s(t)=0$ yields $I_{1}=I_{2}=\tfrac{1}{2}I_{\mathrm{o}}$, while $s(t)=1$ yields $I_{1}=I_{\mathrm{o}}$ and $I_{2}=0$. The resulting coherence-modulated bimode is
\begin{equation}\label{eq:BasicDmatrix}
\mathbf{D}(t)=\tfrac{1}{2}\left(\begin{array}{cc}1+s(t)&0\\0&1-s(t)\end{array}\right),    
\end{equation}
and the total power in the bimode is fixed. In this scenario, the degree of coherence $D(t)$ switches between 0 and 1, corresponding to incoherent and coherent bimodes, respectively. We will make use of coherence modulator~(III) henceforth. The results are applicable to coherence modulator~(II) after noting that $D(t)=\pm1$ rather than $D(t)=0$ and ~1. We present the results for coherence modulator~(I) in Appendix~3.

\subsection{Modal Stokes parameters}

We now introduce the definition of modal SPs for a bimode, which are the real expansion coefficients of the $2\times2$ coherence matrix $\mathbf{G}$ in terms of the usual Pauli matrices:
\begin{equation}\label{eq:GeneralG}
\mathbf{G}=\left(\!\begin{array}{cc}G_{11}&G_{12}\\G_{21}&G_{22}\end{array}\!\right)=\frac{1}{2}\sum_{j=0}^{3}s_{j}\hat{\sigma}_{j}=\frac{1}{2}\left(\!\begin{array}{cc}1+s_{3}&s_{1}-is_{2}\\s_{1}+is_{2}&1-s_{3}\end{array}\!\right),
\end{equation}
where $\{\hat{\sigma}_{j}\}$ are the Pauli matrices:
\begin{equation}\label{eq:Pauli}
\hat{\sigma}_{1}=\left(\begin{array}{cc}0&1\\1&0\end{array}\right),\;
\hat{\sigma}_{2}=\left(\begin{array}{cc}0&-i\\i&0\end{array}\right),\;
\hat{\sigma}_{3}=\left(\begin{array}{cc}1&0\\0&-1\end{array}\right),
\end{equation}
$\hat{\sigma}_{0}=\hat{\mathbb{I}}_{2}$ is the $2\times2$ identity matrix, $s_{0}=\mathrm{Tr}\{\mathbf{G}\}=1$, $\mathrm{Tr}\{\hat{\sigma}_{j}\}=0$ ($j\neq0$), and $\mathrm{Tr}\{\hat{\sigma}_{j}\hat{\sigma}_{k}\}=2\delta_{jk}$ ($j,k=1,2,3$), so that $s_{j}=\mathrm{Tr}\{\hat{\sigma}_{j}\mathbf{G}\}$. The coherence-encoded bimode in Eq.~\ref{eq:Dmatrix} and Eq.~\ref{eq:BasicDmatrix} have the form of Eq.~\ref{eq:GeneralG} with $s_{1}(t)=s_{2}(t)=0$ and $s_{3}(t)=s(t)$ or $s'(t)$.

\subsection{Stokes encoding}

\begin{figure}[t!]
\centering
\includegraphics[width=8.8cm]{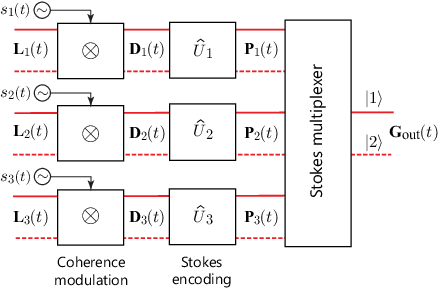}
\caption{Coherence modulation and Stokes encoding. We start with three mutually incoherent bimodes $\mathbf{L}_{j}$, $j=1,2,3$, each of whose degree of coherence is modulated by a data stream $s_{j}(t)$ to yield the coherence-encoded bimodes $\mathbf{D}_{j}(t)$ via coherence modulators followed by Stokes-encoding unitaries $\hat{U}_{j}$ to transform $\mathbf{D}_{j}(t)$ into the Stokes-encoded bimode $\mathbf{P}_{j}(t)$, the three of which are then combined by the Stokes multiplexer into a pair of physical modes underpinning $\mathbf{G}_{\mathrm{out}}(t)$.}
\label{Fig:StokesEncoding}
\end{figure}

The question we pose now is the following: can we construct a \textit{fixed} optical system that modulates the three SPs $\{s_{1},s_{2},s_{3}\}$ for a bimode utilizing three \textit{independent} data streams? We propose the procedure illustrated in Fig.~\ref{Fig:StokesEncoding}. We first prepare three identical bimodes $\mathbf{L}_{j}$ (Eq.~\ref{eq:Lmatrix}), each comprising a pair of mutually uncorrelated modes. However, we add an additional requirement: all three pairs are mutually uncorrelated. This can be realized by preparing 6~mutually uncorrelated modes (see Appendix~1). The need for this additional assumption will be clear shortly. The 6~modes are arranged into three bimodes $\mathbf{L}_{j}$, each of which is directed to a coherence modulator along with a data stream $s_{j}(t)$ to produce \textit{coherence-encoded bimode} $\mathbf{D}_{j}(t)$:
\begin{equation}\label{eq:GenericD}
\mathbf{D}_{j}(t)=\frac{1}{2}\left(\begin{array}{cc}1+s_{j}(t)&0\\0&1-s_{j}(t)\end{array}\right),\;\;j=1,2,3.
\end{equation}

The next step is to transform these bimodes via Stokes-encoding unitaries $\hat{U}_{1}$, $\hat{U}_{2}$, and $\hat{U}_{3}$:
\begin{equation}
\hat{U}_{1}\!=\!\frac{1}{\sqrt{2}}\!\left(\!\!\begin{array}{cc}1&-1\\1&1\end{array}\!\!\right)\!,\;
\hat{U}_{2}\!=\!\frac{1}{\sqrt{2}}\!\left(\!\!\begin{array}{cc}1&i\\i&1\end{array}\!\!\right)\!,\;
\hat{U}_{3}=\left(\!\begin{array}{cc}1&0\\0&1\end{array}\!\right)\!.
\end{equation}
The defining feature of these unitaries is that they are Clifford unitaries that convert Pauli matrices into each other. Specifically, the unitary $\hat{U}_{1}$ converts $\hat{\sigma}_{3}$ into $\hat{\sigma}_{1}$, $\hat{U}_{1}\hat{\sigma}_{3}\hat{U}_{1}^{\dagger}=\hat{\sigma}_{1}$, and the unitary $\hat{U}_{2}$ converts $\hat{\sigma}_{3}$ into $\hat{\sigma}_{2}$, $\hat{U}_{3}\hat{\sigma}_{3}\hat{U}_{3}^{\dagger}=\hat{\sigma}_{2}$, and $\hat{U}_{3}=\hat{\mathbb{I}}_{2}$ of course converts $\hat{\sigma}_{3}$ into itself.

This property of Stokes-encoding unitaries allows them to `move' the diagonal elements in $\mathbf{D}_{j}(t)$ to an off-diagonal position. For example, starting with a diagonal bimode $\mathbf{D}_{j}(t)=\tfrac{1}{2}\hat{\sigma}_{0}+\tfrac{1}{2}s_{j}(t)\hat{\sigma}_{3}$ as given in Eq.~\ref{eq:GenericD}, traversing the Stokes-encoding unitary $\hat{U}_{1}$ yields $\hat{U}_{1}\mathbf{D}_{j}(t)\hat{U}_{j}^{\dagger}=\tfrac{1}{2}\hat{\sigma}_{0}+\tfrac{1}{2}s_{j}(t)\hat{\sigma}_{1}$, thereby moving the data stream $s_{j}(t)$ from the diagonal to the real part of the off-diagonal element. Alternatively, the Stokes-encoding unitary $\hat{U}_{2}$ yields $\hat{U}_{2}\mathbf{D}_{j}(t)\hat{U}_{2}^{\dagger}=\tfrac{1}{2}\hat{\sigma}_{0}+\tfrac{1}{2}s_{j}(t)\hat{\sigma}_{2}$, thereby moving the data stream $s_{j}(t)$ from the diagonal to the imaginary part of the bimode off-diagonal element. Of course, traversing the identity $\hat{U}_{3}$ leaves $s_{j}(t)$ on the diagonal. We therefore direct each of the diagonal coherence-encoded bimodes $\mathbf{D}_{j}(t)$ to a different Stokes-encoding unitary $\hat{U}_{j}$: we direct $\mathbf{D}_{1}(t)$ to $\hat{U}_{1}$, $\mathbf{D}_{2}(t)$ to $\hat{U}_{2}$, and $\mathbf{D}_{3}(t)$ to $\hat{U}_{3}$ [Fig.~\ref{Fig:StokesEncoding}], thus yielding:
\begin{eqnarray}\label{eq:Pmatrices}
\mathbf{P}_{1}(t)\!\!\!\!&=&\!\!\!\!\hat{U}_{1}\mathbf{D}_{1}(t)\hat{U}_{1}^{\dagger}=\frac{1}{2}\left(\begin{array}{cc}1&s_{1}(t)\\s_{1}(t)&1\end{array}\right),\nonumber\\
\mathbf{P}_{2}(t)\!\!\!\!&=&\!\!\!\!\hat{U}_{2}\mathbf{D}_{2}(t)\hat{U}_{2}^{\dagger}=\frac{1}{2}\left(\begin{array}{cc}1&-is_{2}(t)\\is_{2}(t)&1\end{array}\right),\nonumber\\
\mathbf{P}_{3}(t)\!\!\!\!&=&\!\!\!\!\hat{U}_{3}\mathbf{D}_{3}(t)\hat{U}_{3}^{\dagger}=\frac{1}{2}\left(\begin{array}{cc}1+s_{3}(t)&0\\0&1-s_{3}(t)\end{array}\right).
\end{eqnarray}
Each data stream now occupies a position in the coherence matrices corresponding to a different modal SP. We refer to the bimodes $\mathbf{P}_{j}(t)$ as \textit{Stokes-encoded bimodes}. The corresponding equations for coherence modulator~(I) are given in Appendix~3.

\subsection{Stokes multiplexer}

\begin{figure}[t!]
\centering
\includegraphics[width=8.8cm]{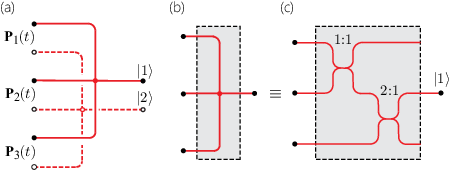}
\caption{(a) A Stokes multiplexer that combines three Stokes-encoded bimodes $\mathbf{P}_{j}(t)$ (Eq.~\ref{eq:Pmatrices}) into a pair of physical modes $|1\rangle$ and $|2\rangle$ that constitute the output field $\mathbf{G}_{\mathrm{out}}(t)$. The Stokes multiplexer comprises two modal combiners. (b) A single modal combiner from (a), which produces either mode $|1\rangle$ or $|2\rangle$. (c) Construction of a modal combiner that combines three modes with equal weights via two couplers with appropriate coupling ratios. Only one of the three outputs is utilized.}
\label{Fig:StokesMultiplexing}
\end{figure}

Finally, we multiplex the three Stokes-encoded bimodes in two physical modes by combining mode-1 from $\mathbf{P}_{1}(t)$, $\mathbf{P}_{2}(t)$, and $\mathbf{P}_{3}(t)$ into mode $|1\rangle$, and combining mode-2 from them into mode $|2\rangle$, as illustrated in Fig.~\ref{Fig:StokesMultiplexing}(a,b). This can be achieved (for either $|1\rangle$ or $|2\rangle$) using a cascade of two couplers or beam splitters with appropriate coupling ratios [Fig.~\ref{Fig:StokesMultiplexing}(c)]. The resulting two-mode field $\mathbf{G}_{\mathrm{out}}(t)$ is then directed to a conventional multiplexer [Fig.~\ref{Fig:Multiplexers}] and thence to the optical channel.

Our goal is for $\mathbf{G}_{\mathrm{out}}(t)$ to be the sum of the Stokes-encoded bimodes, $\mathbf{G}_{\mathrm{out}}(t)=\tfrac{1}{3}\{\mathbf{P}_{1}(t)+\mathbf{P}_{2}(t)+\mathbf{P}_{3}(t)\}$, in which case:
\begin{equation}\label{eq:FullCoherenceMatrix}
\mathbf{G}_{\mathrm{out}}(t)=\frac{1}{6}\left(\begin{array}{cc}3+s_{3}(t)&s_{1}(t)-is_{2}(t)\\s_{1}(t)+is_{3}(t)&3-s_{3}(t)\end{array}\right).
\end{equation}
Summing the coherence matrices for bimodes when combining the corresponding modes \textit{is only valid when the bimodes are uncorrelated}. In general, when combining two bimodes, rather than adding their associated coherence matrices, one should instead form the $4\times4$ coherence matrix that describes the two bimodes \textit{and} the correlation between them. This treatment has been used in maximizing double-slit interference visibility in a vector field \cite{Gori06OL,Gori07OL,Abouraddy17OE}, entropy exchange between binary DoFs \cite{Okoro17Optica,Harling22OE,Harling23JO} and their coherence rank \cite{Harling24PRA,Harling24PRA2,Harling25APLP}, and extending the concept of cross-spectral purity to any binary DoF \cite{Abouraddy26OL}.

Consider combining a pair of bimodes $\mathbf{G}_{a}$ and $\mathbf{G}_{b}$. The associated $4\times4$ coherence matrix $\mathbf{G}$ in block-matrix form is:
\begin{equation}
\mathbf{G}=\frac{1}{2}\left(\begin{array}{cc}\mathbf{G}_{a}&\mathbf{G}_{ab}\\\mathbf{G}_{ba}&\mathbf{G}_{b}\end{array}\right),
\end{equation}
where $\mathbf{G}_{ab}$ and $\mathbf{G}_{ba}$ represent the correlations between the modes underpinning $\mathbf{G}_{a}$ and $\mathbf{G}_{b}$. Combining the two bimodes then corresponds to implementing a $4\times4$ unitary $\hat{U}$ on $\mathbf{G}$ to yield $\mathbf{G}'=\hat{U}\mathbf{G}\hat{U}^{\dagger}=\tfrac{1}{2}\left(\begin{array}{cc}\mathbf{G}_{a}'&\mathbf{G}_{ab}'\\\mathbf{G}_{ba}'&\mathbf{G}_{b}'\end{array}\right)$, after which we restrict ourselves to one output bimode, either $\mathbf{G}_{a}'$ or $\mathbf{G}_{b}'$. To guarantee that this output bimode is simply the sum of the two input bimodes, we need to ensure that the modes underpinning the two bimodes are mutually uncorrelated. We prove this assertion in Appendix~2 (see also \cite{Abouraddy26OL}). Therefore, the sum of $\mathbf{P}_{1}(t)$, $\mathbf{P}_{2}(t)$, and $\mathbf{P}_{3}(t)$ in Eq.~\ref{eq:FullCoherenceMatrix} is valid when  all six modes underpinning the three bimodes are mutually uncorrelated (which can be produced via the scheme described in Appendix~1).

The Stokes multiplexer [Fig.~\ref{Fig:StokesMultiplexing}] is unitary and comprises 6~input and 6~output modes. The coherence matrix $\mathbf{G}_{\mathrm{out}}(t)$ takes up only two of these modes, and we discard the remaining 4~modes. Finally, the constant term along the diagonal of $\mathbf{G}_{\mathrm{out}}(t)$ in Eq.~\ref{eq:FullCoherenceMatrix} is different than that in Eq.~\ref{eq:GeneralG} as a consequence of combining \textit{uncorrelated} bimodes.

A unique feature of S-Comm can be gleaned from Eq.~\ref{eq:FullCoherenceMatrix}. Let us eliminate the third data stream $s_{3}(t)$ by removing the third coherence modulator and Stokes-encoding unitary in Fig.~\ref{Fig:StokesEncoding}. The bimodal field is then:
\begin{equation}
\mathbf{G}_{\mathrm{out}}(t)=\frac{1}{4}\left(\begin{array}{cc}
2&s_{1}(t)-is_{2}(t)\\s_{1}(t)+is_{2}(t)&2
\end{array}\right).
\end{equation}
The diagonal elements here are constant. In other words, recording the modal weights in the optical channel (e.g., by eavesdropping on an optical fiber) reveals no information. Rather, all the data is encoded in correlations between the modes rather than in the modes themselves. This feature of S-Comm provides for a new form of secure communications (see also \cite{Nardi22OL}). 

\subsection{Stokes demultiplexing}

After traversing the optical channel, a conventional multiplexer separates the two physical modes in the input field $\mathbf{G}_{\mathrm{in}}(t)$, which are then directed to a Stokes demultiplexer (the inverse of the Stokes multiplexer). Because of the impact of the optical channel, $\mathbf{G}_{\mathrm{in}}(t)$ is not necessarily equal to $\mathbf{G}_{\mathrm{out}}(t)$ (Eq.~\ref{eq:FullCoherenceMatrix}). We assume here that $\mathbf{G}_{\mathrm{in}}(t)\approx\mathbf{G}_{\mathrm{out}}(t)$ and study the impact of a realistic optical channel on the received coherence matrix elsewhere.

The Stokes demultiplexer for $N_{\mathrm{f}}=2$ is illustrated in Fig.~\ref{Fig:StokesDemultiplexing}(a). Input mode $|1\rangle$ is split equally into three copies, and similarly for mode $|2\rangle$. We thus form three bimodes $\mathbf{Q}_{1}(t)$, $\mathbf{Q}_{2}(t)$, and $\mathbf{Q}_{3}(t)$ by taking one of the copies of mode $|1\rangle$ to correspond to mode-1 in the bimode, and a copy of mode $|2\rangle$ to correspond to mode-2. All three of these bimodes are thus equal to the input field $\mathbf{Q}_{1}(t)=\mathbf{Q}_{2}(t)=\mathbf{Q}_{3}(t)=\mathbf{G}_{\mathrm{in}}(t)$.

\subsection{Stokes decoding}

\begin{figure}[t!]
\centering
\includegraphics[width=8.8cm]{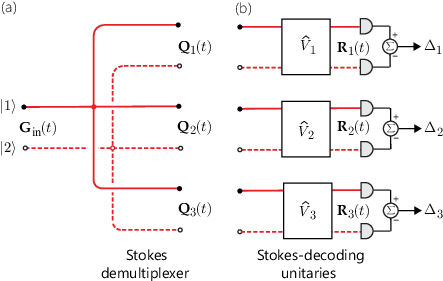}
\caption{(a) Stokes demultiplexer. The modes $|1\rangle$ and $|2\rangle$ in the received bimodal field $\mathbf{G}_{\mathrm{in}}(t)$ are each split into three copies, with a copy from each combined into a bimode, thus yielding three bimodes $\mathbf{Q}_{j}(t)$, $j=1,2,3$. (b) Stokes decoding. Each bimode $\mathbf{Q}_{j}(t)$ traverses a unitary $\hat{V}_{j}=\hat{U}_{j}^{\dagger}$ to yield the bimode $\mathbf{R}_{j}(t)=\hat{V}_{j}(t)\mathbf{Q}_{j}(t)\hat{V}_{j}^{\dagger}$. Differential detection yields the difference $\Delta_{j}(t)$ between the modal weights of $\mathbf{R}_{j}(t)$, from which the data stream $s_{j}(t)$ is recovered.}
\label{Fig:StokesDemultiplexing}
\end{figure}

The goal of Stokes decoding is to selectively move a prescribed data stream in a bimode $\mathbf{Q}_{j}(t)$ to its diagonal. To achieve this, we direct each bimode to a Stokes-decoding unitary $\hat{V}_{j}=\hat{U}_{j}^{\dagger}$:
\begin{equation}
\hat{V}_{1}=\!\frac{1}{\sqrt{2}}\!\left(\!\!\!\begin{array}{cc}1&1\\-1&1\end{array}\!\!\!\right)\!,\;
\hat{V}_{2}\!=\!\frac{1}{\sqrt{2}}\!\left(\!\!\!\begin{array}{cc}1&-i\\-i&1\end{array}\!\!\!\right)\!,\;
\hat{V}_{3}=\left(\!\!\begin{array}{cc}1&0\\0&1\end{array}\!\!\right)\!.
\end{equation}
By directing bimode $\mathbf{Q}_{1}(t)$ to the Stokes-decoding unitary $\hat{V}_{1}$, bimode $\mathbf{Q}_{2}(t)$ to $\hat{V}_{2}$, and $\mathbf{Q}_{3}(t)$ to $\hat{V}_{3}$ [Fig.~\ref{Fig:StokesDemultiplexing}(b)], we obtain the bimodes $\mathbf{R}_{1}(t)$, $\mathbf{R}_{2}(t)$, and $\mathbf{R}_{3}(t)$, respectively:
\begin{eqnarray}
\mathbf{R}_{1}(t)\!\!\!\!&=&\!\!\!\!\!\hat{V}_{1}\mathbf{G}_{\mathrm{in}}(t)\hat{V}_{1}^{\dagger}=\!\frac{1}{6}\!\left(\!\!\!\begin{array}{cc}3+s_{1}(t)&-s_{3}(t)-is_{2}(t)\\-s_{3}(t)+is_{2}(t)&3-s_{1}(t)\end{array}\!\!\!\right)\!,\nonumber\\
\mathbf{R}_{2}(t)\!\!\!\!&=&\!\!\!\!\!\hat{V}_{2}\mathbf{G}_{\mathrm{in}}(t)\hat{V}_{2}^{\dagger}=\!\frac{1}{6}\!\left(\!\!\begin{array}{cc}3+s_{2}(t)&s_{1}(t)+is_{3}(t)\\s_{1}(t)-is_{3}(t)&3-s_{2}(t)\end{array}\!\!\right)\!,\nonumber\\
\mathbf{R}_{3}(t)\!\!\!\!&=&\!\!\!\!\!\hat{V}_{3}\mathbf{G}_{\mathrm{in}}(t)\hat{V}_{3}^{\dagger}=\!\frac{1}{6}\!\left(\!\!\!\begin{array}{cc}3+s_{3}(t)&s_{1}(t)-is_{2}(t)\\s_{1}(t)+is_{2}(t)&3-s_{3}(t)\end{array}\!\!\!\right)\!.
\end{eqnarray}
Now the data streams $s_{1}(t)$, $s_{2}(t)$, and $s_{3}(t)$ are alone along the diagonal of $\mathbf{R}_{1}(t)$, $\mathbf{R}_{2}(t)$, and $\mathbf{R}_{3}(t)$, respectively.

It is important to emphasize that the bimodes $\mathbf{R}_{j}(t)$ do not coincide with the coherence-encoded bimodes $\mathbf{D}_{j}(t)$. However, the diagonal elements of $\mathbf{R}_{j}(t)$ coincide with those of $\mathbf{D}_{j}(t)$, so that detection of their modal weights yield identical outcomes.

\begin{figure*}[t!]
\centering
\includegraphics[width=14.8cm]{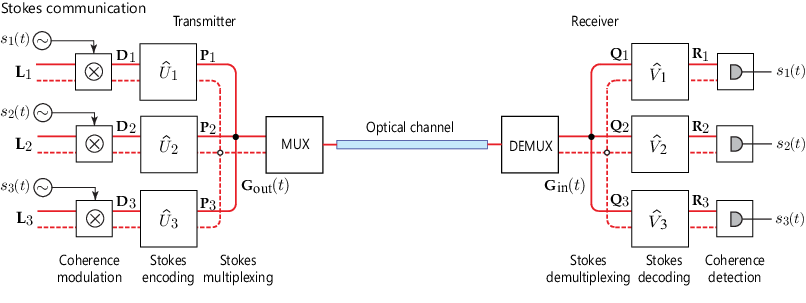}
\caption{Summary of bimodal Stokes communications (S-Comm). In the S-Comm transmitter, each of three identical uncorrelated bimodes $\mathbf{L}_{j}$, $j=1,2,3$, is coherence-modulated by a data stream $s_{j}(t)$ to yield $\mathbf{D}_{j}(t)$; Stokes-encoding unitaries $\hat{U}_{j}$ yield the bimodes $\mathbf{P}_{j}(t)$; which are Stokes-multiplexed into a bimodal field $\mathbf{G}_{\mathrm{out}}(t)$ and then conventionally multiplexed into the optical channel. At the S-Comm receiver, a conventional demultiplexer produces the bimodal field $\mathbf{G}_{\mathrm{in}}(t)$, which is Stokes-demultiplexed into bimodes $\mathbf{Q}_{j}(t)$. Stokes-decoding unitaries $\hat{V}_{j}=\hat{U}_{j}^{\dagger}$ yield the bimodes $\mathbf{R}_{j}(t)$, and the data streams $s_{j}(t)$ are recovered from the difference $\Delta_{j}(t)$ between the modal weights recorded by differential detection.}
\label{Fig:SComm}
\end{figure*}

\subsection{Differential detection}

To recover the data stream $s_{j}(t)$ one directs the bimode $\mathbf{R}_{j}(t)$ to a pair of detectors, the first of which records the modal weight of mode-1, $I_{1}^{(j)}(t)=\tfrac{1}{6}\{3+s_{j}(t)\}$, and the second of which records the modal weight of mode-2, $I_{2}^{(j)}(t)=\tfrac{1}{6}\{3-s_{j}(t)\}$, after which we subtract the modal weights to eliminate the background:
\begin{equation}
\Delta_{j}(t)=I_{1}^{(j)}(t)-I_{2}^{(j)}(t)=\frac{1}{3}s_{j}(t).
\end{equation}
This differential detection scheme eliminates the background even if its value changes over time. Differential detection thus recovers the three data streams $s_{1}(t)$, $s_{2}(t)$, and $s_{3}(t)$ from the bimodes $\mathbf{R}_{1}(t)$, $\mathbf{R}_{2}(t)$, and $\mathbf{R}_{3}(t)$:
\begin{eqnarray}
\Delta_{1}(t)\!\!\!&=&\!\!\!I_{1}^{(1)}-I_{2}^{(1)}\propto s_{1}(t),\nonumber\\
\Delta_{2}(t)\!\!\!&=&\!\!\!I_{1}^{(2)}-I_{2}^{(2)}\propto s_{2}(t),\nonumber\\
\Delta_{3}(t)\!\!\!&=&\!\!\!I_{1}^{(3)}-I_{2}^{(3)}\propto s_{3}(t).
\end{eqnarray}
The protocol for S-Comm is now complete by the recovery of the data streams.

\section{Summary of Stokes communications}

We combine the entire S-Comm system in Fig.~\ref{Fig:SComm}. The starting point is preparing 6~mutually uncorrelated modes arranged into three identical bimodes $\mathbf{L}$. Three data streams $s_{j}(t)$, $j=1,2,3$, modulate the degree of coherence of these bimodes to yield coherence-encoded bimodes $\mathbf{D}_{j}(t)$, which are converted via Stokes-encoding unitaries $\hat{U}_{j}$ to Stokes-encoded bimodes $\mathbf{P}_{j}(t)$, which are and then multiplexed into the physical modes $|1\rangle$ and $|2\rangle$. These two modes are multiplexed into an optical channel; e.g., mapped to the polarization DoF in an SMF, to scalar modes in a pair of SMFs, or to two modes in an MMF, a free-space, or an underwater link. In all of these scenarios, two modes conventionally multiplex two data streams via on-off keying; by contrast, three data streams multiplexed here. Crucially, the number of multiplexed data stream grows quadratically with $N_{\mathrm{f}}$, so that $N_{\mathrm{f}}=3$ modes can multiplex $N_{\mathrm{D}}=8$ data streams.

At the S-Comm receiver, the two physical modes are split into three copies corresponding to the bimodes $\mathbf{Q}_{j}t)$, and Stokes-decoding unitaries $\hat{V}_{j}$ convert them to the bimodes $\mathbf{R}_{j}(t)$, in which the modal SP $s_{j}(t)$ occupies the diagonal. Differential detection then recovers the data streams $s_{j}(t)$, $j=1,2,3$.

\section{Discussion and Future work}

We have outlined here a new form of optical communications (S-Comm) utilizing partially coherent light that promises an increase in channel density over an optical channel -- potentially $\mathcal{O}(N_{\mathrm{f}}^{2})$ data streams multiplexed in $N_{\mathrm{f}}$~physical modes. Much work needs to be done to move this scheme into practice and to clarify its theoretical foundations. Generalizing this scheme to $N_{\mathrm{f}}>2$ requires extending the definition of SPs (Eq.~\ref{eq:GeneralG}) and Pauli matrices (Eq.~\ref{eq:Pauli}) from $N_{\mathrm{f}}=2$ to arbitrary dimensions (see the accompanying paper \cite{Abouraddy26JOSAA2}). We have presented here a full design for the S-Comm scheme. Nevertheless, it appears counter-intuitive that a space spanned by $N_{\mathrm{f}}=2$~modes can be uniquely decomposed into $N_{\mathrm{D}}=3$~independent field components. Similarly, a space spanned by $N_{\mathrm{f}}=3$ modes is separated into $N_{\mathrm{D}}=8$ independent field components.  More work is thus needed to elucidate the mathematical foundation for the particular sense in which the combined bimodes can be uniquely separated without crosstalk, and are thus in some sense orthogonal.

The limitations of S-Comm need to be investigated. For example, the initial optical source from which the bimodes $\mathbf{L}$ are extracted has a finite linewidth $\Delta\lambda$. We shall therefore investigate the spectral efficiency of this scheme. The next limitation is the potential impact of signal-to-noise ratio on the bit error rate. Finally, we need to explore the impact of channel impairments on the transmission. We assumed here that the optical field produced by the transmitter $\mathbf{G}_{\mathrm{out}}(t)$ essentially coincides with that received $\mathbf{G}_{\mathrm{in}}(t)$. However, many deleterious effects in the channel can negatively impair $\mathbf{G}_{\mathrm{in}}(t)$, including relative phases, modal losses, and relative delays incurred between the modes in the optical channel. The quantitative assessment of the impact of these channel impairments on S-Comm will ultimately determine its applicability.

\section{Conclusion}

In conclusion, we have presented a new conception for dense optical communications that makes use of partially coherent light, which we call Stokes communications (S-Comm). We have presented a detailed design for the S-Comm transmitter and receiver for a bimodal field $N_{\mathrm{f}}=2$, whereby $N_{\mathrm{D}}=N_{\mathrm{f}}^{2}-1=3$ data streams are transmitted over an optical channel supporting two modes. We present in an accompanying paper \cite{Abouraddy26JOSAA2} an extension of this description to $N_{\mathrm{f}}>2$, which makes use of the bimodal S-Comm system as a building block. Crucially, the system described here can be implemented in photonic integrated circuit platforms \cite{Capmany20Book,Bogaerts20Nature,Hashemi26arxivTwoMode,Hashemi26LPR3modes,Hashemi26arxiv4Modes,Hashemi26OL,Mor26arxiv}, thereby enabling scalability. S-Comm promises a dramatic increase in channel capacity over an optical channel without increasing the number of modes, and offers a novel modality for secure optical communications.

\section*{Appendix~1: Preparing coherence modes}

\begin{figure}[t!]
\centering
\includegraphics[width=5.5cm]{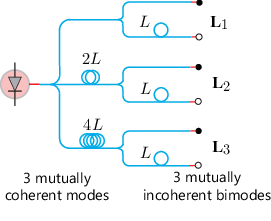}
\caption{Preparing three identical uncorrelated bimodes $\mathbf{L}_{j}$ for bimodal S-Comm. An optical source of finite coherence length $L_{\mathrm{o}}$ is split into three paths with path relative lengths $0,2L$, and $4L$. In a second stage, each path is split into two with relative path lengths 0 and $L$. The total relative lengths in the 6~paths are $0,l,2L,3L,4L$, and $5L$; we assume that $L>L_{\mathrm{o}}$.}
\label{Fig:Source}
\end{figure}

The preparation of three mutually uncorrelated bimodes is depicted in Fig.~\ref{Fig:Source}. Starting with an optical signal (e.g., from a laser diode) of finite linewidth $\Delta\lambda$ and coherence length $L_{\mathrm{o}}$, the field is first split into three paths incorporating relative lengths of $0,2L,4L$, where $L>L_{\mathrm{o}}$, to render them mutually uncorrelated. We next divide each path into two, and introduce a relative length~$L$ between them, thus rendering them mutually uncorrelated bimodes. Moreover, any pair of paths is separated by at least a length~$L$, so that they are all mutually uncorrelated. We arrange the paths into three bimodes, each comprising two paths and described by the coherence matrix $\mathbf{L}=\tfrac{1}{2}\hat{\mathbb{I}}_{2}$. 

\section*{Appendix~2: Combining bimodes}

When two \textit{partially coherent bimodes} $\mathbf{G}_{a}$ and $\mathbf{G}_{b}$ are combined, two new bimodes $\mathbf{G}_{a}'$ and $\mathbf{G}_{b}'$ are produced that need not be weighted sums of $\mathbf{G}_{a}$ and $\mathbf{G}_{b}$. The reason is that $\mathbf{G}_{a}$ and $\mathbf{G}_{b}$ may be correlated \cite{Abouraddy26OL}. The input field is represented by a $4\times4$ coherence matrix $\mathbf{G}$ that includes $\mathbf{G}_{a}$ and $\mathbf{G}_{b}$ and their correlations. We write $\mathbf{G}$ in block-matrix form $\mathbf{G}=\frac{1}{2}\left(\!\!\begin{array}{cc}\mathbf{G}_{a}&\mathbf{G}_{ab}\\
\mathbf{G}_{ba}&\mathbf{G}_{b}
\end{array}\!\!\right)$, where $\mathbf{G}_{a}$ and $\mathbf{G}_{b}$ are superposed with equal weights, $\mathrm{Tr}\{\mathbf{G}\}=1$, and $\mathbf{G}_{ab}$ is a $2\times2$ matrix representing the correlations between the modes in $\mathbf{G}_{a}$ and $\mathbf{G}_{b}$ (but is not a bimode itself) \cite{Abouraddy26OL}. Although $\mathbf{G}_{a}$ and $\mathbf{G}_{b}$ are Hermitian, $\mathbf{G}_{ab}$ and $\mathbf{G}_{ba}$ are \textit{not} necessarily so; nevertheless, $\mathbf{G}_{ab}^{\dagger}=\mathbf{G}_{ab}$. Superposing $\mathbf{G}_{a}$ and $\mathbf{G}_{b}$ at a symmetric coupler corresponds to implementing the following unitary on $\mathbf{G}$:
\begin{equation}
\hat{B}_{\mathrm{b}}=\frac{1}{\sqrt{2}}\left(\!\!\begin{array}{cccc}
1&0&i&0\\
0&1&0&i\\
i&0&1&0\\
0&i&0&1
\end{array}\!\!\right)=
\frac{1}{\sqrt{2}}\!\left(\!\!\begin{array}{cc}1&i\\i&1
\end{array}\!\!\right)\otimes\hat{\mathbb{I}}_{2}=\hat{B}\otimes\hat{\mathbb{I}}_{2},
\end{equation}
yielding a coherence matrix $\mathbf{G}'=\hat{B}_{\mathrm{b}}\mathbf{G}\hat{B}_{\mathrm{b}}^{\dagger}=(\hat{B}\otimes\hat{\mathbb{I}}_{2})\mathbf{G}(\hat{B}^{\dagger}\otimes\hat{\mathbb{I}}_{2})$:
\begin{equation}
\mathbf{G}'=\frac{1}{4}\left(\!\!\!\begin{array}{cc}
\mathbf{G}_{a}+\mathbf{G}_{b}-i\mathbf{G}_{ab}+i\mathbf{G}_{\mathrm{ba}}&
-i\mathbf{G}_{a}+i\mathbf{G}_{b}
+\mathbf{G}_{ab}+\mathbf{G}_{\mathrm{ba}}\\
i\mathbf{G}_{a}-i\mathbf{G}_{b}
+\mathbf{G}_{ab}+\mathbf{G}_{\mathrm{ba}}&
\mathbf{G}_{a}+\mathbf{G}_{b}
+i\mathbf{G}_{ab}-i\mathbf{G}_{\mathrm{ba}}
\end{array}\!\!\!\right).
\end{equation}
If the initial bimodes are mutually uncorrelated, then $\mathbf{G}_{ab}=\mathbf{G}_{ba}=\hat{\mathbf{0}}_{2}$. We further assume that the bimodes initially have equal weights $|\alpha|^{2}=|\beta|^{2}=\tfrac{1}{2}$, so that $\mathbf{G}'$ simplifies to:
\begin{equation}\label{eq:BalancedIncoherentSuperposition}
\mathbf{G}'=\frac{1}{4}\left(\begin{array}{cc}
\mathbf{G}_{a}+\mathbf{G}_{b}&
-i(\mathbf{G}_{a}-\mathbf{G}_{b})\\
i(\mathbf{G}_{a}-\mathbf{G}_{b})&
\mathbf{G}_{a}+\mathbf{G}_{b}
\end{array}\right).
\end{equation}

\begin{figure}[t!]
\centering
\includegraphics[width=6cm]{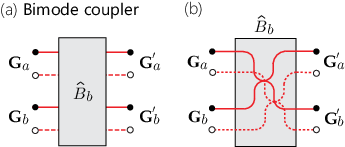}
\caption{(a) A bimode coupler $\hat{B}_{\mathrm{b}}$ that linearly couples bimodes $\mathbf{G}_{a}$ and $\mathbf{G}_{b}$ to yield bimodes $\mathbf{G}_{a}'$ and $\mathbf{G}_{b}'$. (b) The bimode coupler $\mathbf{B}_{\mathrm{b}}$ is constructed of a conventional coupler $\hat{B}$ that combines mode-1 from $\mathbf{G}_{a}$ and $\mathbf{G}_{b}$, and another that combines mode-2 from $\mathbf{G}_{a}$ and $\mathbf{G}_{b}$.}
\label{Fig:CombiningBimodes}
\end{figure}

The coherence matrix $\mathbf{G}'$ describes the entire field emerging from the system in Fig.~\ref{Fig:CombiningBimodes}. The diagonal block matrices in $\mathbf{G}'$ represent the output bimodes $\mathbf{G}_{a}'$ and $\mathbf{G}_{b}'$. If we consider only $\mathbf{G}_{a}'$ and discard $\mathbf{G}_{b}'$, then the new bimode after renormalization is $\mathbf{G}_{a}'=\tfrac{1}{2}\{\mathbf{G}_{a}+\mathbf{G}_{b}\}$, in which the bimodes $\mathbf{G}_{a}$ and $\mathbf{G}_{b}$ are simply added. The discarded bimode $\mathbf{G}_{b}'=\tfrac{1}{2}\{\mathbf{G}_{a}+\mathbf{G}_{b}\}$ has the same structure. However, the bimodes $\mathbf{G}_{a}'$ and $\mathbf{G}_{b}'$ are no longer uncorrelated because $\mathbf{G}'$ in Eq.~\ref{eq:BalancedIncoherentSuperposition} is not block-diagonal. Care must thus be taken subsequently so that all the fields to be combined are mutually uncorrelated, which allows us to continue to add the corresponding bimodes.

Crucially, because the underlying modes are all uncorrelated, even when the bimode is rendered coherent by eliminating one mode, it is nevertheless mutually uncorrelated with the other bimodes, and the corresponding coherence matrices can still be added when the fields are combined.

\section*{Appendix~3: Coherence modulator (I)}

Three mutually uncorrelated but identical bimodes $\mathbf{L}_{j}=\tfrac{1}{2}\hat{\mathbb{I}}_{2}$ are each directed to a coherence modulator~(I) along with a data stream $1-s_{j}(t)$, the logical inverse of $s_{j}(t)$, to produce $\mathbf{D}_{j}(t)=\left(\begin{array}{cc}1&0\\0&1-s_{j}(t)\end{array}\right)$, $j=1,2,3$. The Stokes-encoding unitaries $\hat{U}_{j}$ convert each $\mathbf{D}_{j}(t)$ to yield a Stokes-encoded bimode:
\begin{eqnarray}
\mathbf{P}_{1}(t)&=&\hat{U}_{1}\mathbf{D}_{1}(t)\hat{U}_{1}^{\dagger}\propto\frac{1}{2}\left(\begin{array}{cc}
2-s_{1}(t)&s_{1}(t)\\s_{1}(t)&2-s_{1}(t)
\end{array}\right),\nonumber\\
\mathbf{P}_{2}(t)&=&\hat{U}_{2}\mathbf{D}_{2}(t)\hat{U}_{2}^{\dagger}\propto\frac{1}{2}\left(\begin{array}{cc}
2-s_{2}(t)&-is_{2}(t)\\is_{2}(t)&2-s_{2}(t)
\end{array}\right),\nonumber\\
\mathbf{P}_{3}(t)&=&\hat{U}_{3}\mathbf{D}_{3}(t)\hat{U}_{3}^{\dagger}\propto\left(\begin{array}{cc}1&0\\0&1-s_{3}(t)\end{array}\right).
\end{eqnarray}
Combining these via a Stokes multiplexer yields the field:
\begin{equation}
\mathbf{G}_{\mathrm{out}}(t)\propto\left(\begin{array}{cc}
6-s_{1}(t)-s_{2}(t)&s_{1}(t)-is_{2}(t)\\
s_{1}(t)+is_{2}(t)&6-s_{1}(t)-s_{2}(t)-2s_{3}(t)
\end{array}\right).
\end{equation}

Assuming $\mathbf{G}_{\mathrm{in}}(t)\approx\mathbf{G}_{\mathrm{out}}(t)$ at the receiver, we prepare three copies $\mathbf{Q}_{1}(t)=\mathbf{Q}_{2}(t)=\mathbf{Q}_{3}(t)=\mathbf{G}_{\mathrm{in}}(t)$, each directed to a Stokes-decoding unitary $\hat{V}_{j}$ to yield $\mathbf{R}_{j}(t)=\hat{V}_{j}\mathbf{Q}_{j}(t)\hat{V}_{j}^{\dagger}$:
\begin{equation}
\mathbf{R}_{1}(t)\propto\left(\!\begin{array}{cc}
6-s_{2}(t)-s_{3}(t)&-s_{3}(t)-is_{2}(t)\\
-s_{3}(t)+is_{2}(t)&6-2s_{1}(t)-s_{2}(t)-s_{3}(t)
\end{array}\!\right),\nonumber
\end{equation}
\begin{equation}
\mathbf{R}_{2}(t)\propto\left(\!\begin{array}{cc}
6-s_{1}(t)-s_{3}(t)&s_{1}(t)+is_{3}(t)\\
s_{1}(t)-is_{3}(t)&6-s_{1}(t)-2s_{2}(t)-s_{3}(t)
\end{array}\!\right),\nonumber
\end{equation}
\begin{equation}
\mathbf{R}_{3}(t)\propto\left(\!\begin{array}{cc}
6-s_{1}(t)-s_{2}(t)&s_{1}(t)-is_{2}(t)\\
s_{1}(t)+is_{3}(t)&6-s_{1}(t)-s_{2}(t)-2s_{3}(t)
\end{array}\!\right).
\end{equation}
A pair of detectors record the modal-weight difference $\Delta_{j}(t)$ for each bimode $\mathbf{R}_{j}(t)$, obtaining $\Delta_{j}(t)\propto s_{j}(t)$.

\textbf{Acknowledgments.} This work was funded by the U.S. Office of Naval Research (ONR) under contract N00014-20-1-2789.

\bibliography{diffraction}

@BOOK{SalehBook07,
  AUTHOR =       {B. E. A. Saleh and M. C. Teich},
  TITLE =        {Fundamentals of Photonics},
  PUBLISHER =    {Wiley},
  YEAR =         {2007},
  address =      {},
}

@ARTICLE{Kagalwala13NP,
  AUTHOR =       {K. H. Kagalwala and G. {Di G}iuseppe and A. F. Abouraddy and B. E. A. Saleh},
  TITLE =        {Bell's measure in classical optical coherence},
  JOURNAL =      {Nat. Photon.},
  YEAR =         {2013},
  volume =       {7},
  pages =        {72-78},
}

@ARTICLE{Abouraddy14OL,
  AUTHOR =       {A. F. Abouraddy and K. H. Kagalwala and B. E. A. Saleh},
  TITLE =        {Two-point optical coherency matrix tomography},
  JOURNAL =      {Opt. Lett.},
  YEAR =         {2014},
  volume =       {39},
  pages =        {2411-2414},
}

@ARTICLE{Kagalwala15SR,
  AUTHOR =       {K. H. Kagalwala and H. E. Kondakci and A. F. Abouraddy and B. E. A. Saleh},
  TITLE =        {Optical coherency matrix tomography},
  JOURNAL =      {Sci. Rep.},
  YEAR =         {2015},
  volume =       {5},
  pages =        {15333},
}

@ARTICLE{Okoro17Optica,
  AUTHOR =       {C. Okoro and H. E. Kondakci and A. F. Abouraddy and K. C. Toussaint},
  TITLE =        {Demonstration of an optical-coherence converter},
  JOURNAL =      {Optica},
  YEAR =         {2017},
  volume =       {4},
  pages =        {1052-1058},
}

@ARTICLE{Waller12NP,
  AUTHOR =       {L. Waller and G. Situ and J. W. Fleischer},
  TITLE =        {Phase-space measurement and coherence
synthesis of optical beams},
  JOURNAL =      {Nat. Photon.},
  YEAR =         {2012},
  volume =       {6},
  pages =        {474-479},
}

@ARTICLE{Nardi22OL,
  AUTHOR =       {A. Nardi and S. Divitt and M. Rossi and F. Tebbenjohanns and A. Militaru and M. Frimmer and L. Novotny},
  TITLE =        {Encoding information in the mutual coherence of spatially separated light beams},
  JOURNAL =      {Opt. Lett.},
  YEAR =         {2022},
  volume =       {47},
  pages =        {4588-4591},
}

@ARTICLE{Abouraddy17OE,
  AUTHOR =       {A. F. Abouraddy},
  TITLE =        {What is the maximum attainable visibility by a partially coherent electromagnetic field in {Y}oung's double-slit interference?},
  JOURNAL =      {Opt. Express},
  YEAR =         {2017},
  volume =       {25},
  pages =        {18320-18331},
}

@ARTICLE{Harling24PRA,
  title = {Locked entropy in partially coherent optical fields},
  author = {M. Harling and V. A. Kelkar and K. C. Toussaint and A. F. Abouraddy},
  journal = {Phys. Rev. A},
  volume = {109},
  pages = {L021501},
  year = {2024},
}

@article{Harling24PRA2,
  title = {Isoentropic partially coherent optical fields that cannot be interconverted unitarily},
  author = {M. Harling and V. A. Kelkar and K. C. Toussaint and A. F. Abouraddy},
  journal = {Phys. Rev. A},
  volume = {110},
  issue = {1},
  pages = {013505},
  numpages = {11},
  year = {2024},
  month = {Jul},
  publisher = {American Physical Society},
}

@ARTICLE{Gori06OL,
  AUTHOR =       {F. Gori and M. Santarsiero and R. Borghi},
  TITLE =        {Vector mode analysis of a {Y}oung interferometer},
  JOURNAL =      {Opt. Lett.},
  YEAR =         {2006},
  volume =       {31},
  pages =        {858-860},
}

@ARTICLE{Gori07OL,
  AUTHOR =       {F. Gori and M. Santarsiero and R. Borghi},
  TITLE =        {Maximizing {Y}oung's fringe visibility through reversible optical transformations},
  JOURNAL =      {Opt. Lett.},
  YEAR =         {2007},
  volume =       {32},
  pages =        {588-590},
}

@article{roques2024Light,
  title={Measuring, processing, and generating partially coherent light with self-configuring optics},
  author={C. Roques-Carmes and S. Fan and D. Miller},
  journal={Light Sci. Appl.},
  year={2024},
  volume =       {13},
  pages =        {260},
}

@ARTICLE{Bogaerts20Nature,
  AUTHOR =       {W. Bogaerts and D. P{\'e}rez and J. Capmany and D. A. B. Miller and J. Poon and D. Englund and F. Morichetti and A. Melloni},
  TITLE =        {Programmable photonic circuits},
  JOURNAL =      {Nature},
  YEAR =         {2020},
  volume =       {586},
  pages =        {207–216},
}

@ARTICLE{Willner15AOP,
  AUTHOR =       {A. E. Willner and H. Huang and Y. Yan and Y. Ren and N. Ahmed and G. Xie and C. Bao and L. Li and Y. Cao and Z. Zhao and J. Wang and M. P. J. Lavery and M. Tur and S. Ramachandran and A. F. Molisch and N. Ashrafi and S. Ashrafi},
  TITLE =        {Optical communications using orbital angular momentum beams},
  JOURNAL =      {Adv. Opt. Photon.},
  YEAR =         {2015},
  volume =       {7},
  pages =        {66-106},
}

@ARTICLE{Li14AOP,
  AUTHOR =       {G. Li and N. Bai and N. Zhao and C. Xia},
  TITLE =        {Space-division multiplexing: the next frontier in optical communication},
  JOURNAL =      {Adv. Opt. Photon.},
  YEAR =         {2014},
  volume =       {6},
  pages =        {413-487},
}

@article{Harling22OE,
  title={Reversible inter-degree-of-freedom optical-coherence conversion via entropy swapping},
  author={M. Harling and V. Kelkar and C. Okoro and M. Diouf and A. F. Abouraddy and K. C. Toussaint},
  journal={Opt. Express},
  volume={30},
  pages={29584-29597},
  year={2022},
}

@ARTICLE{Harling23JO,
  AUTHOR =       {M. Harling and V. Kelkar and A. F. Abouraddy and K. C. Toussaint},
  TITLE =        {Reversible coherence conversion across optical degrees-of-freedom: a tutorial},
  JOURNAL =      {J. Opt.},
  YEAR =         {2023},
  volume =       {25},
  pages =        {053502},
}

@ARTICLE{Harling25APLP,
  AUTHOR =       {M. Harling and C. Stevenson and K. C. Toussaint and A. F. Abouraddy},
  TITLE =        {Optical communications through highly scattering channels using the coherence-rank},
  JOURNAL =      {APL Photonics},
  YEAR =         {2025},
  volume =       {10},
  pages =        {076116},
}

@book{Saleh25book,
  AUTHOR =       {B. E. A. Saleh},
  TITLE =        {Quantum Photonics: Bimodes, Qubits, and Biphotons},
  YEAR =         {2025},
  publisher =    {Springer},
  address =      {},
}

@ARTICLE{Miller25Optica,
  AUTHOR =       {D. A. B. Miller and C. Roques-Carmes and C. G. Valdez and A. R. Kroo and M. Vlk and S. Fan and O. Solgaard},
  TITLE =        {Universal programmable and self-configuring optical filter},
  JOURNAL =      {Optica},
  YEAR =         {2025},
  volume =       {12},
  pages =        {1417-1426},
}

@ARTICLE{Dong24Nature,
  AUTHOR =       {B. Dong and F. Br{\"u}ckerhoff-Pl{\"u}ckelmann and L. Meyer and J. Dijkstra and I. Bente and D. Wendland and A. Varri and S. Aggarwal and N. Farmakidis and M. Wang and G. Yang and J. S. Lee and Y. He and E. Gooskens and D.-L. Kwong and P. Bienstman and W. H. P. Pernice and H. Bhaskaran},
  TITLE =        {Partial coherence enhances parallelized photonic computing},
  JOURNAL =      {Nature},
  YEAR =         {2024},
  volume =       {632},
  pages =        {55-62},
}

@ARTICLE{Liu25LPR,
  AUTHOR =       {X. Liu and X. Li and S. A. Ponomarenko and F. Wang and X. Peng and Y. Cai and C. Liang},
  TITLE =        {Unlocking secure optical multiplexing with spatially incoherent light},
  JOURNAL =      {Laser Photon. Rev.},
  YEAR =         {2025},
  volume =       {19},
  pages =        {2401534},
}

@ARTICLE{Peng21P,
  AUTHOR =       {D. Peng and Z. Huang and Y. Liu and Y. Chen and F. Wang and S. A. Ponomarenko and Y. Cai},
  TITLE =        {Optical coherence encryption with structured random light},
  JOURNAL =      {PhotoniX},
  YEAR =         {2021},
  volume =       {2},
  pages =        {6},
}

@Book{Capmany20Book,
  AUTHOR =       {J. Capmany and D. P{\'e}rez},
  TITLE =        {Programmable Integrated Photonics},
  Publisher =    {Oxford Univ. Press},
  YEAR =         {2020},
  address =      {Oxford},
}

@ARTICLE{Abouraddy26AOP,
  AUTHOR =       {A. F. Abouraddy and B. E. A. Saleh},
  TITLE =        {Structured coherence: {A} modern perspective on optical coherence as a resource},
  JOURNAL =      {arXiv:2608.06356},
  YEAR =         {2026},
  volume =       {},
  pages =        {},
}

@ARTICLE{Abouraddy26OPN,
  AUTHOR =       {A. F. Abouraddy and M. Harling and A. Shiri and K. C. Toussaint},
  TITLE =        {Taking advantage of structured coherence},
  JOURNAL =      {Opt. Photon. News},
  YEAR =         {2026},
  volume =       {37},
  issue =        {7},
  pages =        {26-33},
}

@ARTICLE{Brooks85JLT,
  AUTHOR =       {J. L. Brooks and R. H. Wentworth and R. C. Youngquist and M. Tur and B. Y. Kim and H. J. Shaw},
  TITLE =        {Coherence multiplexing of fiber-optic interferometric sensors},
  JOURNAL =      {J. Lightwave Technol.},
  YEAR =         {1985},
  volume =       {3},
  pages =        {1062-1072},
}

@ARTICLE{Griffin95JLT,
  AUTHOR =       {R. A. Griffin and D. D. Sampson and D. A. Jackson},
  TITLE =        {Coherence coding for photonic code-division multiple access networks},
  JOURNAL =      {J. Lightwave Technol.},
  YEAR =         {1995},
  volume =       {13},
  pages =        {1826-1837},
}

@ARTICLE{Sampson97FiberIO,
  AUTHOR =       {D. D. Sampson and G. J. Pendock and R. A. Griffin},
  TITLE =        {Photonic code-division multiple-access communications},
  JOURNAL =      {Fiber Integ. Opt.},
  YEAR =         {1997},
  volume =       {16},
  pages =        {129-157},
}

@ARTICLE{Kats04IEEETWC,
  AUTHOR =       {G. Kats and S. Arnon},
  TITLE =        {Analysis of optical coherence multiplexing networks for satellite communication},
  JOURNAL =      {IEEE Trans. Wireless Comm.},
  YEAR =         {2004},
  volume =       {3},
  pages =        {1444-1451},
}

@ARTICLE{Meijerink07IEEEJSTQE,
  AUTHOR =       {A. Meijerink and R. O. Taniman and G. H. L. M. Heideman and W. {van E}tten},
  TITLE =        {Coherence multiplex system topologies},
  JOURNAL =      {IEEE J. Sel. Top. Quant. Electron.},
  YEAR =         {2007},
  volume =       {13},
  pages =        {1433-1445},
}

@ARTICLE{Richardson13NP,
  AUTHOR =       {D. J. Richardson and J. M. Fini and L. E. Nelson},
  TITLE =        {Space-division multiplexing in optical fibers},
  JOURNAL =      {Nat. Photon.},
  YEAR =         {2013},
  volume =       {7},
  pages =        {354-362},
}

@ARTICLE{Hashemi26arxivTwoMode,
  AUTHOR =       {A. Hashemi and A. Shiri and B. E. A. Saleh and A. Blanco-Redondo and A. F. Abouraddy},
  TITLE =        {Programmable on-chip synthesis and reconstruction of partially coherent two-mode optical fields},
  JOURNAL =      {APL Photon.},
  YEAR =         {2026},
  volume =       {11},
  pages =        {076112},
}

@ARTICLE{Hashemi26arxiv4Modes,
  AUTHOR =       {A. Hashemi and A. Shiri and B. E. A. Saleh and A. Blanco-Redondo and A. F. Abouraddy},
  TITLE =        {On-chip control of the coherence matrix of four-mode partially coherent light: rank, entropy, and modal Stokes parameters},
  JOURNAL =      {arXiv:2601.18797},
  YEAR =         {2026},
  volume =       {},
  pages =        {},
}

@ARTICLE{Hashemi26OL,
  AUTHOR =       {A. Hashemi and A. Shiri and B. E. A. Saleh and A. Blanco-Redondo and A. F. Abouraddy},
  TITLE =        {Single-shot on-chip measurement of the modal Stokes parameters},
  JOURNAL =      {Opt. Lett.},
  YEAR =         {2026},
  volume =       {51},
  pages =        {3489-3492},
}

@ARTICLE{Hashemi26LPR3modes,
  AUTHOR =       {A. Hashemi and A. Shiri and B. E. A. Saleh and A. Blanco-Redondo and A. F. Abouraddy},
  TITLE =        {On-chip measurement of the {S}tokes-{G}ell-{M}ann parameters for three-mode partially coherent light},
  JOURNAL =      {arXiv:2606.12233},
  YEAR =         {2026},
  volume =       {},
  pages =        {},
}

@ARTICLE{Mor26arxiv,
  AUTHOR =       {P.-A. Mor and A. R. Kroo and C. G. Valdez and M. {\v S}imi{\'c} and A. Karnieli and G. Cavicchioli and Z. Sun and V. Grimaldi and S. Fan and O. Solgaard and D. A. B. Miller and C. Roques-Carmes},
  TITLE =        {Separating partially coherent light},
  JOURNAL =      {	arXiv:2603.15517},
  YEAR =         {2026},
  volume =       {},
  pages =        {},
}

@ARTICLE{Liu23PRAppl,
  AUTHOR =       {X. Liu and C. Liang and Y. Cai and S. A. Ponomarenko},
  TITLE =        {Axial correlation revivals and number factorization with structured random waves},
  JOURNAL =      {Phys. Rev. Appl.},
  YEAR =         {2023},
  volume =       {20},
  pages =        {L021004},
}

@ARTICLE{Baghdady16OE,
  AUTHOR =       {J. Baghdady and K. Miller and K. Morgan and M. Byrd and S. Osler and R. Ragusa and W. Li and B. M. Cochenour and E. G. Johnson},
  TITLE =        {Multi-gigabit/s underwater optical communication link using orbital angular momentum multiplexing},
  JOURNAL =      {Opt. Express},
  YEAR =         {2016},
  volume =       {24},
  pages =        {9794-9805},
}

@ARTICLE{Davenport21OME,
  AUTHOR =       {J. J. Davenport and M. Diab and P. J. Deka and A. Tripathi and K. Madhav and M. M. Roth},
  TITLE =        {Photonic lanterns: a practical guide to filament tapering},
  JOURNAL =      {Opt. Mater. Express},
  YEAR =         {2021},
  volume =       {11},
  pages =        {2639-2649},
}

@ARTICLE{LeonSaval10OE,
  AUTHOR =       {S. G. Leon-Saval and A. Argyros and J. Bland-Hawthorn},
  TITLE =        {Photonic lanterns: a study of light propagation in multimode to single-mode converters},
  JOURNAL =      {Opt. Express},
  YEAR =         {2010},
  volume =       {18},
  pages =        {8430-8439},
}

@ARTICLE{Martinez-Becerril24OE,
  AUTHOR =       {A. C. Martinez-Becerril and S. Luo and L. Li and J. T. R. Pag{\'e} and L. Giner and R. A. Abrahao and J. S. Lundeen},
  TITLE =        {Reconfigurable unitary transformations of optical beam arrays},
  JOURNAL =      {Opt. Express},
  YEAR =         {2024},
  volume =       {32},
  pages =        {41111-41126},
}

@ARTICLE{Abouraddy26JOSAA2,
  AUTHOR =       {A. F. Abouraddy and B. E. A. Saleh},
  TITLE =        {Towards dense optical communications with partially coherent light: {II. M}ultimode {S}tokes communications},
  JOURNAL =      {accompanying paper},
  YEAR =         {2026},
  volume =       {},
  pages =        {},
}

@ARTICLE{Abouraddy26OL,
  AUTHOR =       {A. F. Abouraddy and B. E. A. Saleh},
  TITLE =        {Optical cross-purity},
  JOURNAL =      {Opt. Lett.},
  YEAR =         {2026},
  volume =       {51},
  pages =        {833-836},
}

@ARTICLE{Ip07JLT,
  AUTHOR =       {E. Ip and J. M. Kahn},
  TITLE =        {Digital equalization of chromatic dispersion and polarization mode dispersion},
  JOURNAL =      {J. Lightwave Technol.},
  YEAR =         {2007},
  volume =       {25},
  pages =        {2033-2043},
}

@ARTICLE{Brackett90IEEEJSAC,
  AUTHOR =       {C. A. Brackett},
  TITLE =        {Dense wavelength division multiplexing networks: {P}rinciples and applications},
  JOURNAL =      {IEEE J. Sel. Areas Commun.},
  YEAR =         {1990},
  volume =       {8},
  pages =        {948-964},
}

@Book{Agrawal21Book,
  AUTHOR =       {G. P. Agrawal},
  TITLE =        {Fiber-Optic Communication Systems},
  Publisher =    {John Wiley \& Sons},
  YEAR =         {2021},
  address =      {Hoboken, New Jersey},
}

@incollection{Winzer13chapter,
  author      = "P. J. Winzer and R. Ryf and S. Randel",
  title       = "Spatial multiplexing using multiple-input multiple-output signal processing",
  editor      = "I. P. Kaminow and T. Li and A. E. Willner",
  booktitle   = "Optical Fiber Telecommunications {VI-B}: Systems and Networks",
  publisher   = "Academic Press/Elsevier",
  address     = "",
  year        = 2013,
  pages       = "",
}

\end{document}